# High Throughput Production of Transparent Conductive Single-Walled Carbon Nanotube Films *via* Advanced Floating Catalyst Chemical Vapor Deposition

Qiang Zhang[1,2], Weiya. Zhou[1-4,*], Kewei Li[1,2], Nan Zhang[1,2], Yanchun Wang[1,3], Zhuojian Xiao[1,2], Qingxia Fan[1,2], Sishen Xie[1-4,*]

[1]Beijing National Laboratory for Condensed Matter Physics, Institute of Physics, Chinese Academy of Sciences, Beijing 100190, China

[2]University of Chinese Academy of Sciences, Beijing 100049, China

[3]Beijing Key Laboratory for Advanced Functional Materials and Structure Research, Beijing 100190, China

[4]Songshan Materials Laboratory, Dongguan, Guangdong 523808, China

*To whom correspondence should be addressed. E-mail: wyzhou@iphy.ac.cn, ssxie@iphy.ac.cn

Single-walled carbon nanotube (SWCNT) films are promising materials for transparent conductive films (TCFs) with potential applications in flexible displays, touch screens, solar cells and solid-state lighting[1,2]. However, further reductions in resistivity and in cost of SWCNT films are necessary for high quality TCF products[3]. Here, we report an improved floating catalyst chemical vapor deposition method to directly and continuously produce ultrathin and freestanding SWCNT films at the hundred meter-scale. Both carbon conversion efficiency and SWCNT TCF yield are increased by three orders of magnitude relative to the conventional floating catalyst chemical vapor deposition. After doping, the film manifests a sheet resistance of 40 ohm/sq at 90% transmittance, representing record performance for large-scale SWCNT films. Our work provides a new avenue to accelerate the industrialization of SWCNT films as TCFs.

Transparent conductive films (TCFs) have pervaded modern technologies since they represent an essential component of various optoelectronic devices. Currently, indium tin oxide (ITO) is the most widely used transparent conductive material[1]. The brittle nature and the limited resource of indium, however, present many challenges for ITO applications in flexible electronics. Therefore, alternative transparent conductive materials, e.g., carbon nanotubes (CNTs)[4], graphene[5], metal nanowires[6], and metal meshes[7], have received extensive attention. Among these promising candidates, SWCNT TCFs exhibit great potential because of their outstanding electrical, optical, and mechanical properties, good flexibility, and high environmental stability[8]. The main methods for production of transparent SWCNT films include liquid-phase processing and floating catalyst chemical vapor deposition (FCCVD)[3]. Although solution-based film fabrication is readily scalable and low-cost, the CNT dispersion process can inevitably lead to contamination and shortening of the nanotubes, thus reducing the performance of a film. Gas filtration and then press transfer are the main procedures for FCCVD-based film fabrication in FCCVD [4]. Gas filtration allows good control of CNT synthesis, but produces a low-yield and still requires multiple steps. Inspired by the CNT fiber synthesis via FCCVD[9,10], we have developed an advanced FCCVD to directly and continuously produce a high-quality, freestanding, and transparent conductive SWCNT film at a large scale.

The freestanding SWCNT films were produced in a vertical FCCVD reactor. Feedstock for CNT synthesis and nitrogen carrier gas are introduced continuously from the top of a reactor tube. Tubular thin CNT films are continuously exported from the outlet of the reactor. The produce speed of the tubular film is 50 ~ 600 m/h. To demonstrate the scalability of this method, two sets of equipment (#1 and #2) with different scales were built as shown in Fig 1. Subsequent discussion is based on the large-scale equipment #2 with a diameter of 10 cm, unless otherwise specified.

Thin CNT film fabrication by our method is of ultra-high yield and high carbon conversion efficiency (the percentage of input carbon converted to CNTs). The yield of CNT film can be over 50 $m_2$/h and the carbon conversion efficiency can be up to 25%. Both the yield and the carbon conversion efficiency here are improved by over three orders of magnitude compared to those in reported works[11–17], as shown in Fig. 2a.

The electrical properties of SWCNT films over a period of 90 days under ambient conditions demonstrate a high stability (Fig. 2b). For a transmittance (Tr) of 90%, the pristine CNT film shows a sheet resistance ($R_s$) of 180 $\Omega$/sq and its conductivity can be improved 3 - 4-fold with $HNO_3$ or gold chloride doping. The best transparent conductivity observed was ~ 40 $\Omega$/sq in $R_s$ at 90% Tr by $HNO_3$ doping, and the $R_s$ deteriorated to 65 $\Omega$/sq in 2 days. The change of $R_s$ in the first few days results from spontaneous doping (the pristine films) and de-doping ($HNO_3$ doping) in air. With $AuCl_3$ doping, the $R_s$ was around 50 $\Omega$/sq at 90% Tr, and remained at this level even after 3 months storage in ambient conditions. Variation in $R_s$ versus Tr of SWCNT films with $AuCl_3$ doping, and comparisons of these values with those of previously reported high-performance large-scale CNT films (over 100 $cm_2$) from gas filtration[14,16], liquid-phase processing[18–20] and dry-drawing from CNT forests[21], are shown in Fig. 2c. So, relative to the samples reported so far in the literature, the stable $R_s$ of 50 $\Omega$/sq at 90% Tr of the synthesized film represents the record performance for large-scale CNT films.

The excellent optoelectronic performance which determines whether the SWCNT films could be applied as TCFs (Fig. 2) is primarily ascribed to the high-quality SWCNTs and the unique microstructures of the film. Optical absorption spectra (Fig. 2d) indicate that the film is composed of SWCNTs with a mean diameter of ~2.2 nm and good crystallinity. According to scanning electron microscopy (SEM) and transmission electron microscopy (TEM) images (Fig 2e,), the SWCNT films exhibit a continuously porous and reticulate microstructure with numerous Y-type junctions. These Y-type junctions are formed at high growth temperature and have longer interbundle connections, which benefits carrier transport[11,22]. In addition, we also propose that the ultra-long SWCNTs that comprise the continuous transparent films also contribute to high conductivity[23]. It is worth mentioning that the transmittance of freestanding CNT films ranges from 97% to 60% and can be directly adjusted during film synthesis. The freestanding films of different transmittances mounted on plexiglass frames or polyethylene terephthalate (PET) manifest their optical homogeneity (Fig 2g).

It is noted that less than 10% of SWCNTs are individual in the synthesized films, while the mean bundle diameter is around 18 nm. These bundles are quite straight and homogeneous, only rarely forming loops or circles. This is significantly different from those in the films prepared by other methods[9,12,15]. CNT agglomeration in both the liquid and the gas phase usually causes inhomogeneous bundle structures such as loops, which decrease the electrical property of CNT films[3,8].

How to collect and store such a thin CNT film, especially one with transmittance > 80%, is a big challenge. A setup has been designed for direct and continuous collection of the freestanding film (Fig. 3). Transparent electrodes can be obtained directly by using a transparent substrate such as PET or polyethylene. Some low-surface-energy materials are also good flexible substrates, such as tracing papers. A ~50 m long roll and ~4.5 m sheet of transparent film on tracing paper are shown in Fig. 3c. The magnified images in the insets of Fig. 3c illustrate that the as-collected films are transparent and homogeneous. Especially, the as-collected CNT film can be directly peeled off as a freestanding film again, or can be transferred to some other target substrates (Fig 3d).

In summary, we have shown an advanced FCCVD technique to directly produce a large-area homogeneous SWCNT film with excellent transparent conductivity, which has considerable potential for applications as TCFs in flexible electronics, touch sceen, photoresponsive sensors, energy devices, and so forth[3,24]. To further improve film performance with the $R_s$ less than 10 Ω/sq at 90% Tr, which would meet the requirements of all current applications of TCFs[3,8], attention should be focused on both CNT synthesis[15,24] (in particular the enrichment of a special type of CNTs) and film collection[25]. The high carbon conversion efficiency and stable controllability in composition and excellent performance of the as-synthesized large-scale films indicate that this method will be of great significance for the economical and high throughput production of CNT films, especially as TCFs.

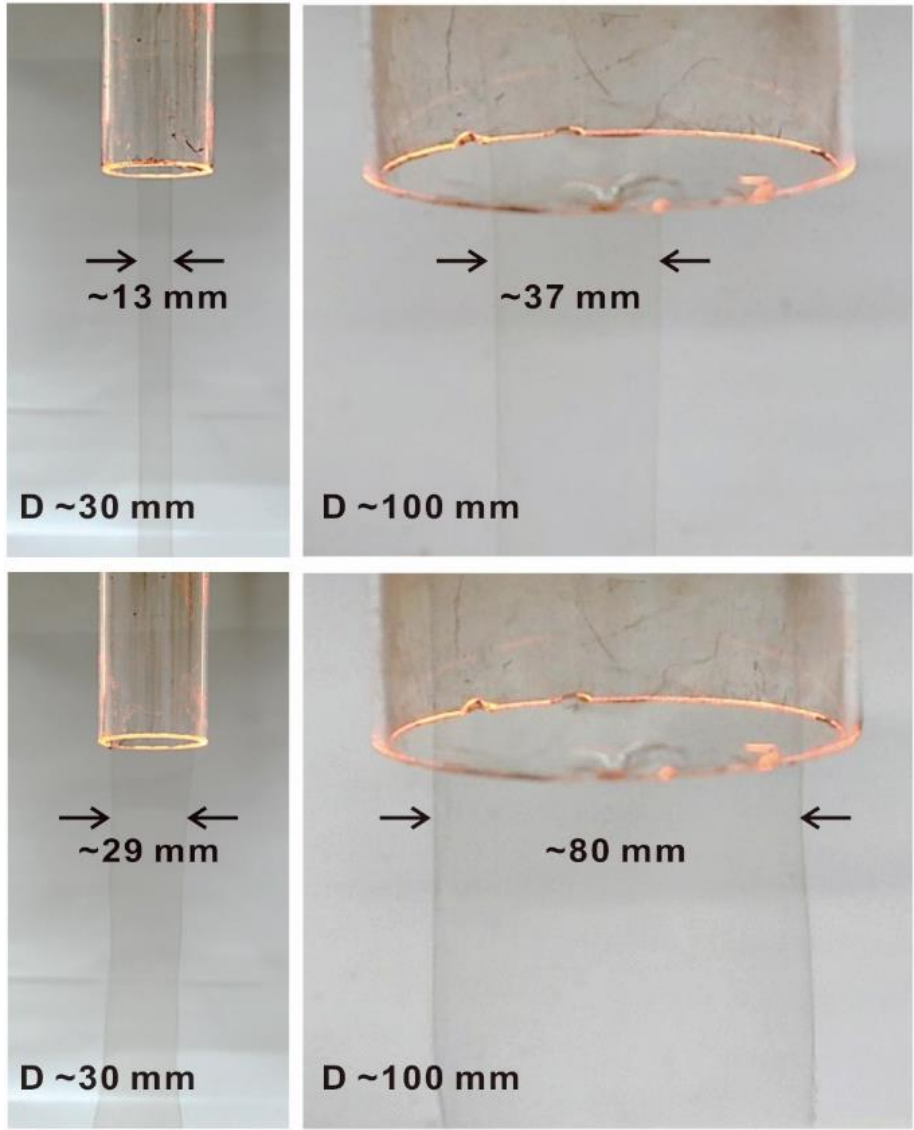

**Figure 1 | Optical images of the freestanding tubular CNT films with different diameters.**
**Left**: the films exported from Equipment #1 (the synthesis tube diameter ~30 mm). **Right**: the films exported from Equipment #2 (the synthesis tube diameter ~100 mm).

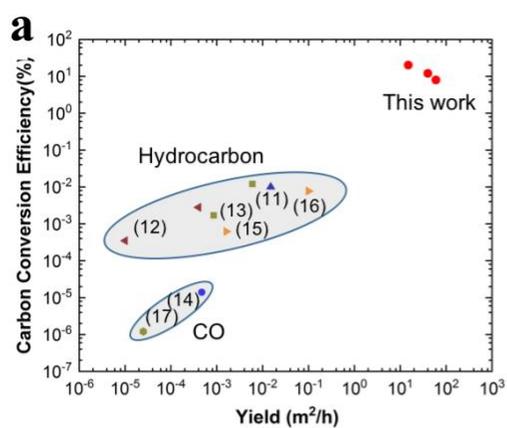

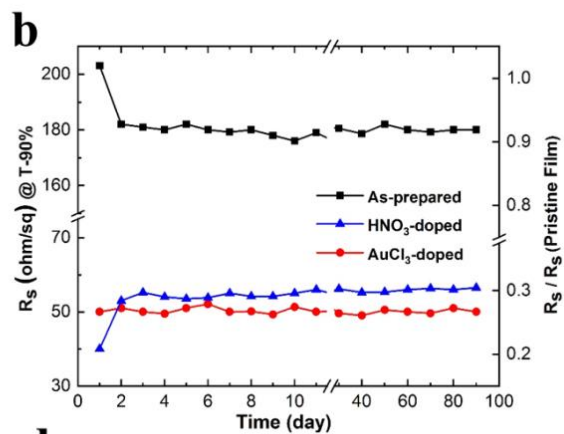

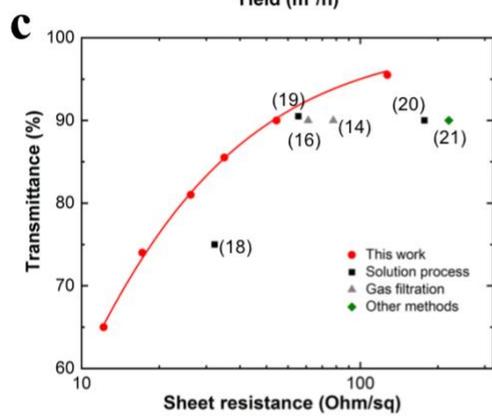

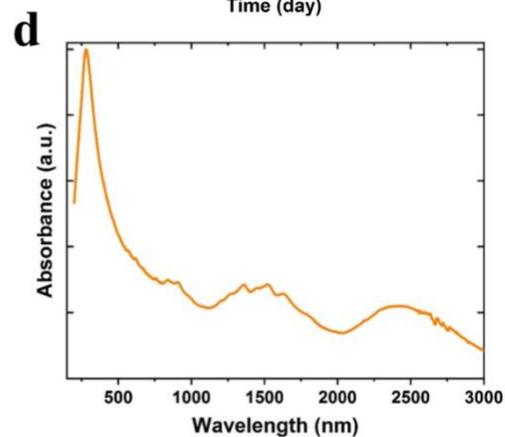

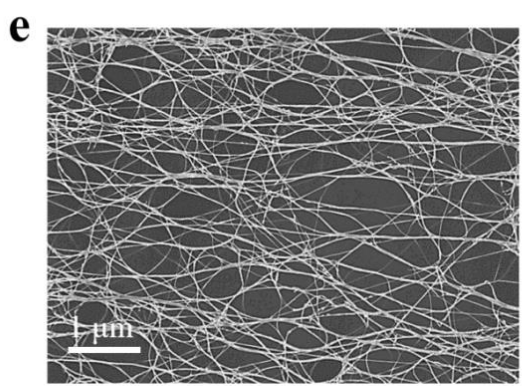

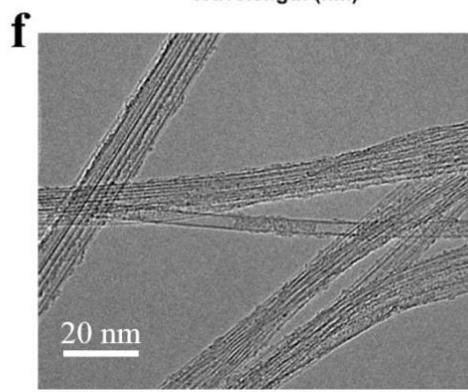

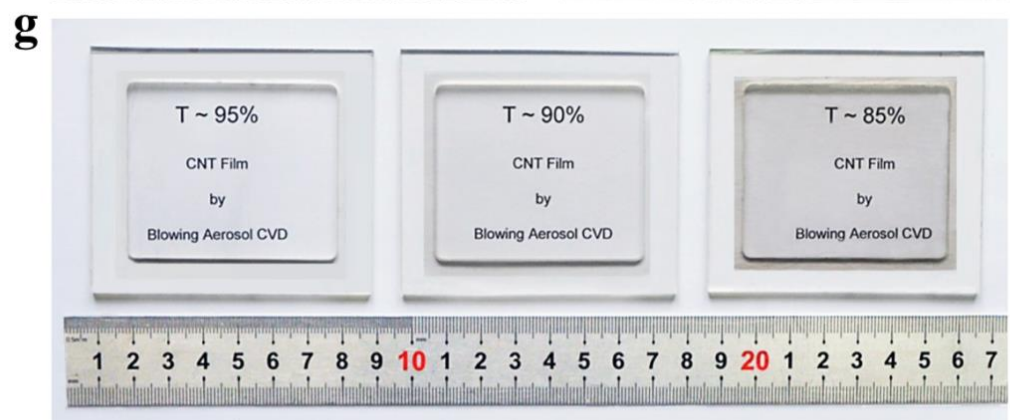

**Figure 2 | Characterization of SWCNT films as TCFs. (a)** Transparent CNT film yield and carbon conversion efficiency in this work, together with previously reported FCCVD results based on hydrocarbon (refs 11-13,15,16) and carbon monoxide (refs 14,17) as the carbon source. **(b)** Variation in and comparisons of $R_s$ of pristine, $AuCl_3$ doped and $HNO_3$ doped SWCNT films under ambient conditions. Film conductivity is improved 3 - 4-fold after $AuCl_3$ or $HNO_3$ doping. $R_s$ changes in the first few days owing to the spontaneous doping and de-doping of CNT films, then becomes extremely stable (relative change in resistance < 5%). **(c)** Sheet resistance versus transmittance at 500 nm of $AuCl_3$ doped films. The best reported performances of large-area CNT films (over 100 cm2) produced by other methods, i.e. gas filtration based on FCCVD (refs 14, 16), a liquid process (refs 18-20) and dry-drawing from multi-walled CNT forests (ref. 21), are shown for comparison. (d) Optical absorption spectra of the pristine as-synthesized film. The film consists of SWCNTs with a mean diameter of around 2.2 nm. **(e, f)** Typical SEM images of the SWCNT synthesized films with 95% and 90% Tr, showing the continuous SWCNT bundle network and robust Y-type junctions formed at the growth temperature. The CNT bundles are straight and homogeneous, only rarely forming loops. **(g)** Plexiglass frames covered by optically homogeneous and freestanding films with various transmittances.

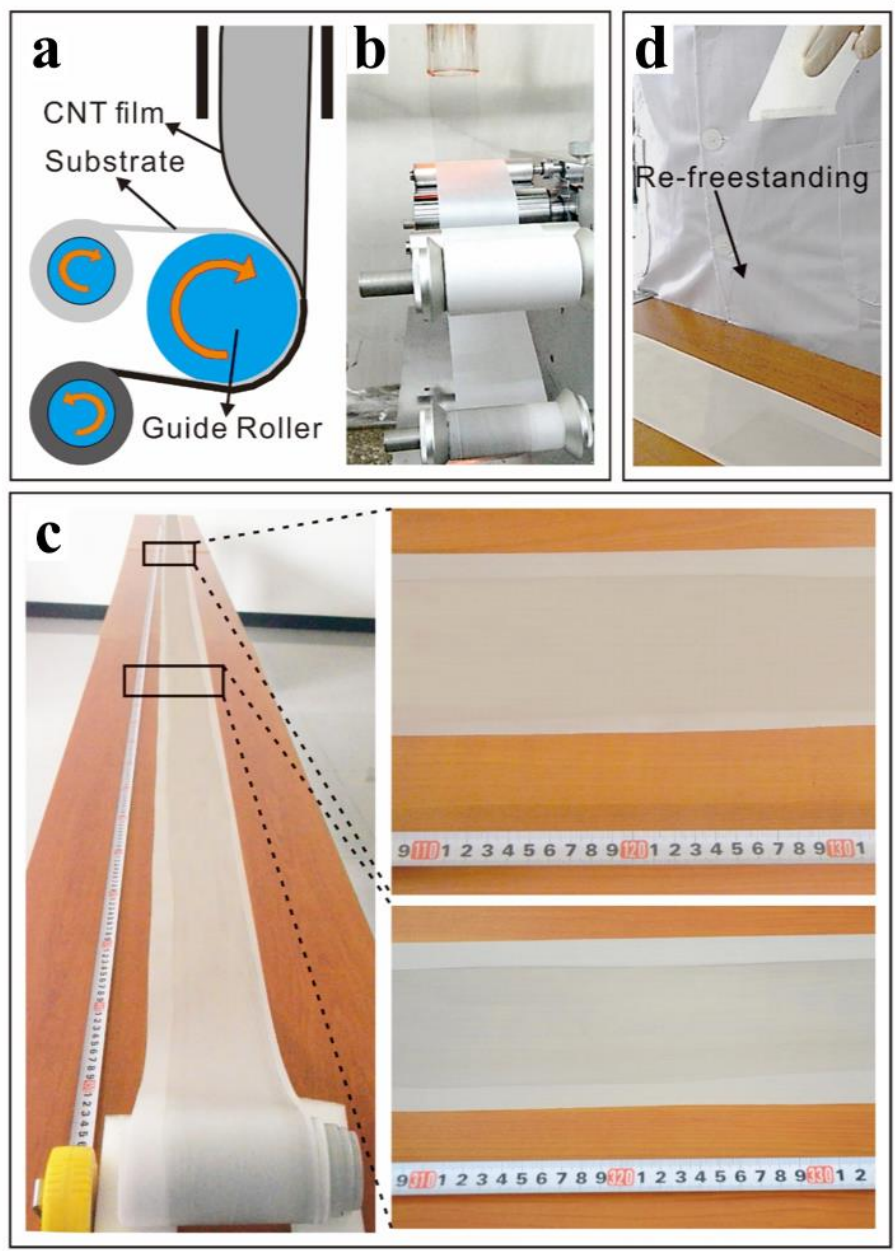

**Figure 3 | The continuous collection of a tubular film. (a)** Schematic of a set-up for continuous film collection on a flexible substrate. During collection, the tubular film is collapsed and attached to a flexible substrate over a guide roller. The composite structure composed of a CNT film and substrate is then collected on a winder. **(b)** Photograph of the film collection on tracing paper at ~ 150 m/h for Equipment #2. **(c)** Photograph of a 4.5 m sheet and 50 m roll of uniform SWCNT film (~ 10 cm width) on tracing paper and magnified photographs of the framed regions. **(d)** Freestanding film re-obtained by directly peeling off the tracing paper.

# References

1.  Ellmer, K. Past achievements and future challenges in the development of optically transparent electrodes. *Nat. Photonics* **6**, 809–817 (2012).

2.  Yu, L., Shearer, C. & Shapter, J. Recent Development of Carbon Nanotube Transparent Conductive Films. *Chem. Rev.* **116**, 13413–13453 (2016).

3.  Hu, L., Hecht, D. S. & Grüner, G. Carbon Nanotube Thin Films: Fabrication, Properties, and Applications. *Chem. Rev.* **110**, 5790–5844 (2010).

4.  Nasibulin, A. G. *et al.* Multifunctional free-standing single-walled carbon nanotube films. *ACS Nano* **5**, 3214–3221 (2011).

5.  Bae, S. *et al.* Roll-to-roll production of 30-inch graphene films for transparent electrodes. *Nat. Nanotechnol.* **5**, 574–578 (2010).

6.  Hu, L. *et al.* Scalable Coating and Properties of Transparent, Flexible, Silver Nanowire Electrodes. *ACS Nano* **4**, 2955–2963 (2010).

7.  Wu, H. *et al.* Electrospun metal nanofiber webs as high-performance transparent electrode. *Nano Lett.* **10**, 4242–4248 (2010).

8.  Zhang, Q., Wei, N., Laiho, P. & Kauppinen, E. I. Recent Developments in Single-Walled Carbon Nanotube Thin Films Fabricated by Dry Floating Catalyst Chemical Vapor Deposition. *Topics in Current Chemistry* (2017). doi:10.1007/s41061-017-0178-8

9.  Li, Y.L., Kinloch, I. A. & Windle, A. H. Direct Spinning of Carbon Nanotube Fibers from Chemical Vapor Deposition Synthesis. *Science (80-. ).* **304**, 276–278 (2004).

10. Nasibulin, A. G. *et al.* A novel hybrid carbon material. *Nat. Nanotechnol.* **2**, 156–161 (2007).

11. Ma, W. *et al.* Directly synthesized strong, highly conducting, transparent single-walled carbon nanotube films. *Nano Lett.* **7**, 2307–2311 (2007).

12. Hussain, A. *et al.* Floating catalyst CVD synthesis of single walled carbon nanotubes from ethylene for high performance transparent electrodes. *Nanoscale* **10**, 9752–9759 (2018).

13. Ding, E. X. *et al.* Highly conductive and transparent single-walled carbon nanotube thin films from ethanol by floating catalyst chemical vapor deposition. *Nanoscale* **9**, 17601–17609 (2017).

14. Liao, Y. *et al.* Tuning Geometry of SWCNTs by CO 2 in Floating Catalyst CVD for High-Performance Transparent Conductive Films. *Adv. Mater. Interfaces* **5**, 1–10 (2018).

15. Cheng, H. M. *et al.* Ultrahigh-performance transparent conductive films of carbon-welded isolated single-wall carbon nanotubes. *Sci. Adv.* **4**, eaap9264 (2018).

16. Wang, B. W. *et al.* Continuous Fabrication of Meter-Scale Single-Wall Carbon Nanotube Films

and their Use in Flexible and Transparent Integrated Circuits. *Adv. Mater.* **30**, 1–8 (2018).

17. Mustonen, K. *et al.* Gas phase synthesis of non-bundled, small diameter single-walled carbon nanotubes with near-armchair chiralities. *Appl. Phys. Lett.* **107**, (2015).

18. Wu, Z. *et al.* Transparent, conductive carbon nanotube films. *Science (80-. ).* **305**, 1273–6 (2004).

19. Dan, B., Irvin, G. C. & Pasquali, M. Continuous and scalable fabrication of transparent conducting carbon nanotube films. *ACS Nano* **3**, 835–843 (2009).

20. Geng, H.-Z. *et al.* Effect of Acid Treatment on Carbon Nanotube-Based Flexible Transparent Conducting Films. *J. Am. Chem. Soc.* **129**, 7758–7759 (2007).

21. Feng, C. *et al.* Flexible, stretchable, transparent conducting films made from superaligned carbon nanotubes. *Adv. Funct. Mater.* **20**, 885–891 (2010).

22. Sun, D. *et al.* Flexible high-performance carbon nanotube integrated circuits. *Nat. Nanotechnol.* **6**, 156–161 (2011).

23. Behabtu, N. *et al.* Strong, light, multifunctional fibers of carbon nanotubes with ultrahigh conductivity. *Science (80-. ).* **339**, 182–6 (2013).

24. De Volder, M. F. L., Tawfick, S. H., Baughman, R. H. & Hart, A. J. Carbon nanotubes: present and future commercial applications. *Science (80-. ).* **339**, 535–539 (2013).

25. Koziol, K. *et al.* High-performance carbon nanotube fiber. *Science (80-. ).* **318**, 1892–1895 (2007).

## Acknowledgements

This work was financially supported by the National Key R&D Program of China (Grant No. 2018YFA0208402), the National Natural Science Foundation of China (11634014, 51172271 and 51372269), the "Strategic Priority Research Program" of the Chinese Academy of Sciences (XDA09040202), the National Basic Research Program of China (Grant No. 2012CB932302).